\documentclass[useAMS,usenatbib]{mn2e}

\title[ion-proton pulsars]{Ion-proton pulsars}
\author[P. B. Jones]{P. B. Jones\thanks{E-mail:
p.jones1@physics.ox.ac.uk}  \\
University of Oxford, Department of Physics, Denys Wilkinson Building,\\
Keble Road, Oxford OX1 3RH, U.K.}

\begin{document}

\date{}

\pagerange{\pageref{firstpage}--\pageref{lastpage}} \pubyear{}

\maketitle

\label{firstpage}

\begin{abstract}

Evidence derived with minimal assumptions from existing published observations is presented to show that an ion-proton plasma is the source of radio-frequency emission in millisecond and in normal isolated pulsars.  There is no primary involvement of electron-positron pairs.  This conclusion has also been reached by studies of the plasma composition based on well-established particle-physics processes in neutron stars with positive  polar-cap corotational charge density.  This work has been published in a series of papers which are also summarized here. It is now confirmed by simple analyses of the observed radio-frequency characteristics, and its implications for the further study of neutron stars are outlined.

\end{abstract}


\section{Introduction}

We believe that a strong case has been developed in a series of publications (Jones 2010a - 2016) for an ion-proton plasma as the source of emission in radio-loud pulsars.  The initial approach was to determine the plasma composition in the open magnetosphere above the polar caps in ${\bf \Omega}\cdot{\bf B} < 0$ neutron stars from well-established particle-physics processes. (Here ${\bf \Omega}$ is the rotational angular velocity and ${\bf B}$ the polar-cap magnetic flux density: the polar-cap Goldreich-Julian charge density is positive.)  This lead immediately to the recognition that the plasma must have a significant proton component arising from photo-nuclear reactions in the electromagnetic showers produced by electrons that are accelerated towards the neutron-star surface.  Either pair creation (if present) or photoelectric ionization of accelerated ions are sources of these electrons.  Later it was realized that the reverse electron flux from photoelectric ionization limits the electric field component antiparallel with ${\bf B}$ as would pair creation were it present. Photoelectric ionization transition rates are rapidly increasing functions of ion Lorentz factor so that particle acceleration is a self-limiting process resulting in moderately relativistic but not ultra-relativistic energies.  Pair creation by single-photon magnetic conversion is not normally possible above the polar cap, but the presence of ions and protons with only moderate Lorentz factors leads to strong growth of longitudinal and quasi-longitudinal Langmuir modes. (In this paper, the term Langmuir mode is used to describe any longitudinal or quasi-longitudinal unstable mode in a system of two or more parallel particle beams of different velocities and of any charged particle type.)

The complete process is not particularly simple and, unfortunately, depends quite naturally on two parameters that are not well-known, the surface ion atomic number $Z$ and the neutron star surface temperature $T_{s}$, more precisely the temperature of the surface area of about one steradian centred on the polar cap, which is the source of the photo-ionizing black-body photons.  An ion-proton plasma is formed for any $Z$ such that ions at the point of leaving the surface retain some bound electrons provided also that values of $T_{s}$ are such that the black-body radiation field has a large enough photon density to produce adequate photoelectric transition rates.  Therefore, 
the formation of an ion-proton plasma is possible for quite wide intervals of these parameters, in particular, temperatures $T_{s} > 1-2\times 10^{5}$ K are adequate. Even so, there may be doubts that suitable values would be present in cooling neutron stars to account for the number of isolated old radio-loud pulsars with ages as large as $10^{2-3}$ Myr.  For this reason, observational evidence that the emission region contains ions and protons but few if any electrons and positrons is very desirable provided it can be interpreted with minimal and uncontentious assumptions.

It was not recognized until most of the work had been completed (Jones 2016)
that such evidence already existed in the pulse-longitude dependence of the integrated profiles of circularly-polarized intensity, the Stokes parameter $V$, exemplified particularly in the high-resolution and good signal-to-noise ratio measurements of Karastergiou \& Johnston (2006) for a set of 17 pulsars.  Section 2 of the present paper contains a survey of this, and of other less direct but still persuasive observational evidence for an ion-proton plasma.  
It also includes a review of estimates of the emission region altitude, which is of prime importance in the interpretation of the circular polarization data.

Early papers (Jones 2010a - 2013a) converge towards this paper's view of the ion-proton plasma.  Thus some results given in those papers have either been superseded or are of only marginal relevance.   In particular, it was recognized only in Jones (2014c) that the presence of a density of electrons or positrons small compared with the Goldreich-Julian density can reduce to zero the Langmuir-mode growth rate which would in its absence exist in an ion-proton plasma. The significance of this result is that it provides a specific mechanism for the bi-stability associated with the phenomena of mode-changes and nulls.  To correct this, Section 3 gives a summary of the properties of the ion-proton plasma including the case of a small background electron-positron component which is likely to be relevant to an understanding of mode-changes, nulls and subpulse drift.  The implications, for both signs of ${\bf \Omega}\cdot{\bf B}$, of the ion-proton plasma in radio-loud pulsars are considered in Section 4. In particular, we believe it shows that the open magnetosphere current, which relates the spin-down torque at the neutron-star surface to that beyond  the light cylinder, is set by conditions at the neutron-star surface rather than the light cylinder.

\section{Observational evidence}

The characteristics of what we describe as the primary radio-frequency emission are almost universal.  Excluding a small number of {\it sui generis} cases, spectral indices are large and negative.  At low frequencies, in cases where observations exist, there is usually a turn-over below $\sim 100$ MHz.   Luminosities have been estimated by Szary et al (2014) for all pulsars which have been observed at $1.4$ GHz, solely from the flux densities at that frequency.  The luminosity distribution is more than two orders of magnitude wide but, remarkably, it is independent of position in the $P -\dot{P}$ plane.  It has also been noted many times (Jenet et al 1998; Kramer et al 1999; Espinoza et al 2013) that the spectra of radio-loud millisecond pulsars (MSP) are broadly similar to those of normal pulsars.

Thus spectral properties appear to have no significant trends of variation over five orders of magnitude in $B$, and three orders of magnitude in rotational period $P$.  It might be thought that this conclusion in itself does not seem obviously consistent with the canonical model of a secondary electron-positron plasma source.  But there are more specific concerns.

\subsection{Radio-frequency power generation}

In relation to the emission process, we believe that the radio-frequency energy generated per unit charge accelerated at the polar cap, equal to $W = L/f_{GJ}$, where $L$ is the luminosity and $f_{GJ}$ is the Goldreich-Julian rate at which unit charges leave the polar cap, is an interesting parameter.  The luminosities were estimated by Jones (2014b) for a set of 29 pulsars which appeared in the spectral compilation published by Malofeev et al (1994) and also had a sufficient signal-to-noise ratio to be listed in the paper of Weltevrede, Edwards \& Stappers (2006).  The luminosities found there are typically one or two orders of magnitude smaller than the mean value of $\sim 10^{29}$ erg s$^{-1}$ given by Szary et al.  Values of $W$ are distributed over more than two orders of magnitude.  In a few cases, for example, $29$ GeV for B1642-03, they can be large enough to suggest possibly the presence of a caustic in one of the variables defining the beam.  But the average value, $W = 4.3$ GeV, is also large.

How is it possible that this value could be generated by either the secondary electron-positron pairs produced by a single ultra-relativistic electron or positron or by unit charge of an ion-proton plasma?  For the electron-positron case the relativistic Penrose condition (see Buschauer \& Benford 1977) must be satisfied. The optimum electron-positron energy distribution which does so is given by 
$N_{\pm}(\delta(\gamma - \gamma_{1}) + \delta(\gamma - \gamma_{2}))$ per primary electron in the neutron-star frame, where $\gamma$ is the electron or positron Lorentz factor and the constants are usually assumed to be of the order of $\gamma_{1,2} \sim 10^{1-2}$. Here $N_{\pm}$ is the secondary-pair multiplicity.  Given the absence of any accelerating field in the emission region, application of the conservation laws for energy and for the momentum component parallel with ${\bf B}$ shows that the maximum energy in photons that can be extracted is,
\begin{eqnarray}
W = N_{\pm}mc^{2}\frac{(\gamma_{1} - \gamma_{2})^{2}}{(\gamma_{1} + \gamma_{2})},
\end{eqnarray}
emphasizing the fact that it is the velocity difference between the beams which makes radio-frequency energy generation possible. (This excludes the generation of curvature radiation.) For $W = 4.3$ GeV, a dense plasma with large $N_{\pm} \sim 10^{2}$ would be necessary for the values of $\gamma_{1,2}$ which are usually assumed and are necessary for adequate Langmuir-mode growth rates.  But there is no evidence that such a favourable distribution forms or that if formed, it would have a similar density in pulsars in all parts of the $P - \dot{P}$ plane.  Our conclusion is that, even in this most favourable case, the electron-positron plasma is unlikely to produce the values of $W$ that are observed.

Owing to the differing charge-to-mass ratio of ions and protons, the $\delta$-function velocity distribution occurs naturally in that plasma.  Application of the kinematic conservation laws to this case, specifically equation (14) of Jones (2014b), shows that, owing to the baryonic masses of the particles, the values of $W$ estimated from the observed luminosities can be achieved in an ion-proton plasma.

Our use of energy and momentum conservation in the emission region would be invalid if there remained any acceleration field ${\bf E}_{\parallel}$.  The effect of radio-frequency energy generation is to reduce the Penrose-condition velocity difference, but a residual ${\bf E}_{\parallel}$ would have the reverse effect in the case of an ion-proton plasma owing to the differing charge-to-mass ratio of the components.  This would, in principle, allow greater generation of radio-frequency energy than the conservation-law values given by Jones (2014b) and might be the explanation for the few very large values of $W$ that are seen.

\subsection{The emission altitude}

Estimates of absolute radio-emission altitudes are always model-dependent.  It is not appropriate to describe here the various assumptions that different authors have made, but we refer to the papers cited in Table 1.  Estimates are usually given in the form of upper limits: these can be superseded by smaller upper limits.  Thus the papers cited have been limited to more recent work except where actual values, with or without errors, have been given.  The results are a miscellany and comparisons for a given pulsar are not possible except in the cases of B0329+54 and B1133+16.  Absolute radii $r_{e}$ are given in the right-hand column.  Errors, where given, are large.  Noutsos et al (2015) attempted to obtain $r_{e}$ for 11 pulsars but found in 7 cases that their procedure gave  values either negative or consistent with zero. They do not give details for these cases but it is possible that their inclusion would have been relevant to the estimation of an average $r_{e}$ for the whole set. Considering all the values given in Table 1 our conclusion is that it is not possible to give an average value for $r_{e}$.

The size of the emission region $\Delta r_{e}$ can be estimated with fewer assumptions and less model-dependence.  Here, the results are all upper limits.  Kramer et al (1997) and Hassall et al (2012) show by using profiles at different frequencies that, for normal pulsars with periods $P\sim 1$ s, the emission regions are compact. They place upper limits on the effects of aberration and retardation, and make no assumptions about radius-to-frequency mapping. (Also, for a substantial set of MSP, Kramer et al 1999 were able to quote the extremely small value $\Delta r_{e} < 2.4$ km as an upper limit to the transit-time difference between radiation at the different observed frequencies.)

The general similarity of normal pulsar and MSP radio spectra is consistent with their being produced in similar plasma conditions.  Goldreich-Julian particle number densities are $\propto BP^{-1}$ and at a given radius are usually smaller in MSP than in normal pulsars but by no more than an order of magnitude, so that plasma conditions are not too dissimilar.  A typical MSP light-cylinder radius is only $R_{LC}\sim 200$ km.  Therefore, an emission radius of perhaps one half of this would require a very compact $\Delta r_{e}$.

On balance, and owing to the similarity of plasma density in MSP and in normal pulsars, we accept the LOFAR results of Hassall et al.  Their general validity will be assumed in Sections 2.3 and 2.4.

\begin{table}
\caption{The Table summarizes the estimates of absolute emission radius $r_{e}$ and the size of the emission region $\Delta r_{e}$ given in some recent publications.  They are for individual pulsars except in the work of von Hoensbroech \& Xilouris.  The superscripts denote as follows: $a$ core components; $b$ conal components; $c$ 610 MHz; $d$ 1408 MHz.}

\begin{tabular}{@{}llrr@{}}
\hline       
    Authors  &  Pulsars     &  $\Delta r_{e}$     &   $r_{e}$                      \\
\hline
                    &          &  km    &  km      \\
\hline                                                  \\                  
von Hoensbroech     &        &     &                   \\
 \& Xilouris, 1997  &      &  -  & $210\pm 180^{a}$  \\
                                   &        &     & $420\pm200^{b}$   \\
                                                                  \\
Kramer et al, 1997                 & 0329+54&  $<320$ &  -  \\
                                   & 0355+54&  $<50$  &  -  \\
                                   & 0540+23&  $<120$ &  -  \\
                                   & 1133+16&  $<310$ &  -  \\
                                   & 1706-16&  $<210$ &  -  \\
                                   & 1929+10&  $<110$ &  -  \\
                                   & 2020+28&  $<160$ &  -  \\
                                   & 2021+51&  $<360$ &  -  \\
                                                                \\
Thomas \& Gangadhara, 2010         & 1839+09&  -  &  $50^{a}$   \\
                                   &        &        &  $60^{b}$   \\
                                   & 1916+14&  -  &  $100^{a}$   \\
                                   &        &  -  &  $300^{b}$   \\
                                   & 2111+46&  -  &  $500^{c}$   \\
                                   &        &  -  &  $80^{d}$   \\
                                                               \\
Karuppusamy, Stappers              &        &     &              \\
\& Serylak, 2011                   & 1133+16&  -  &  $  560$    \\                              
                                                                  \\
Hassall et al, 2012                & 0329+54& $<128$ & $<183$    \\
                                   & 0809+74& $<384$ &  -    \\
                                   & 1133+16& $<59$  & $<110$   \\
                                   & 1919+21& $<49$  &  -     \\
                                                                \\
Noutsos et al, 2015                & 0823+26&  -  & $144^{+136}_{-134}$    \\             
                                   & 0834+06&  -  & $452^{+437}_{-432}$  \\
                                   & 1133+16&  -  & $349^{+158}_{-150}$   \\
                                   & 1953+50&  -  & $177^{+74}_{-119}$    \\

\hline
\end{tabular}

\end{table}

\subsection{Frequency spectrum}

Pulsar radio spectra in cases where sufficient low-frequency observations exist have a turnover, usually below $100$ MHz, and above that a power-law decrease in intensity with large negative spectral index.  There may be a break, a change of index, usually below $1$ GHz and to more rapid intensity decrease.  Almost all the power is emitted well below $1$ GHz.  Some information about the distribution of index values can be found for those pulsars which have fluxes at $400$ and $1400$ MHz listed in the ATNF catalogue (Manchester et al 2005).  This shows that the spectral index is almost uncorrelated with the parameter $X = B_{12}P^{-1.6}$, which is a measure of the capacity of a neutron star to produce pairs above the polar cap in the ${\bf \Omega}\cdot{\bf B} > 0$ case, derived from Fig. 1 of Harding \& Muslimov (2002).

The emission region can be regarded only as a black box within which plasma turbulence exists.  The radiation frequency is well below the threshold for cyclotron absorption and the only natural frequency present in the box is the rest-frame plasma frequency $\omega^{c}_{p}$.  It is natural to attempt to relate this to the observed spectrum as in Jones (2013b). In the case of a secondary electron-positron plasma, the rest-frame plasma frequency is,
\begin{eqnarray}
\omega^{c}_{p} = \left(\frac{8\pi n_{GJ}N_{\pm}{\rm e}^{2}}{m\gamma_{e}}\right)^{1/2}
\end{eqnarray}
where $n_{GJ}$ is the number density of the primary beam of electrons or positrons and $\gamma_{e}$ the average secondary electron-positron Lorentz factor.  It determines the typical wavenumber of the fluctuations from whose growth turbulence develops.  The transfer of energy to higher rather than lower wavenumbers is a general property of developing turbulence, leading in the example of incompressible fluids to a power-law spectrum with the $-5/3$ Kolmogorov index. Recent work by Zrake \& East (2016) on force-free magnetic turbulence, though not immediately applicable here, has produced further evidence for the ubiquity of the Kolmogorov law. Therefore, it is natural here to associate $\omega^{c}_{p}$ with the low-frequency turnover region of the spectrum at a typical radiation frequency,
\begin{eqnarray}
\nu_{obs} = \gamma_{e}\omega^{c}_{p}/\pi \approx 25\left(N_{\pm}\gamma_{e}\frac{B_{12}}{P}\right)^{1/2} \left(\frac{R}{r_{e}}\right)^{3/2}               {\rm GHz},
\end{eqnarray}
where $R$ is the neutron-star radius and $B_{12}$ the polar-cap surface magnetic flux density in units of $10^{12}$ G. This is in order of magnitude disagreement with observation for acceptable values of $N_{\pm}\gamma_{e}$ and for values of $r_{e}$ favoured by Table 1. In our view, this represents a serious problem for the canonical electron-positron source model.  

For an ion-proton plasma, equation (3) is replaced by,
\begin{eqnarray}
\nu_{obs} =     4.2\times 10^{2}\left(\gamma_{A,Z}\frac{ZB_{12}}{AP}\right)^{1/2}\left(\frac{R}{r_{e}}\right)^{3/2} \hspace{5mm}        {\rm MHz}
\end{eqnarray}
in which $\gamma_{A,Z}$ is the ion Lorentz factor in the neutron-star frame.  Here, a turn-over below $100$ MHz is easily possible. This conclusion was reached in Jones (2013b) specifically in the case of B1133+16.

\subsection{Circular polarization}

Circular polarization provides the most definitive evidence for an ion-proton plasma.  In Jones (2016) we have referred to the paper of Karastergiou \& Johnston (2006) as a consequence of the number of pulsar integrated profiles observed, the signal-to-noise ration and the time resolution in that work.  But the crucial feature of the circular polarization can also be seen in the papers of Han et al (1998) and of Yan et al (2011) and Dai et al (2015) on MSP.

The assumptions made here are minimal.  We assume as in Section 2.3 the existence of an emission region above the polar cap whose upper boundary could be defined roughly as the surface of last absorption for radio-frequency radiation.  Above this, the radiation propagates through plasma moving outward from the polar cap.  Observationally, we know that the radiation generally has strong linear polarization: this must be a feature of the emission process but it is modified by propagation through the plasma at $r > r_{e}$.  As it leaves the emission region, it propagates as a linear combination of the normal modes of the plasma whose polarizations for wavevector ${\bf k}$ are parallel with ${\bf k}\times({\bf k }\times {\bf B})$ and ${\bf k}\times {\bf B}$ for the O- and E-modes, respectively.  The E-mode refractive index is negligibly different from unity, but the O-mode refractive index differs from unity at angular frequency $\omega$ by
\begin{eqnarray}
\Delta n_{O} = - \sum_{i}\frac{\omega^{2}_{p}\sin^{2}\theta_{k}}
{2\gamma^{3}_{i}\tilde{\omega}^{2}} \approx -\sum_{i}\frac{2\omega^{2}_{p}\theta^{2}_{k}}
{\gamma^{3}_{i}\omega^{2}(\theta^{2}_{k} + \gamma^{-2}_{i})^{2}},
\end{eqnarray}
in which $\tilde{\omega} = \omega - k_{z}v^{i}_{z}$ and $\omega^{2}_{p} =
4\pi n_{i}q^{2}_{i}/m_{i}$, where $q_{i}$ is the charge of particle type $i$. The local direction of ${\bf B}$ is the $z$-axis, $\theta_{k}$ is the angle between ${\bf k}$ and ${\bf B}$, $n_{i}$ and $\gamma_{i}$ are the particle number density and Lorentz factor. Equation (5) is valid for a particular particle type, provided $\gamma_{i}\tilde{\omega} \ll \omega^{i}_{B}$
(the radiation angular frequency in the particle rest frame is small compared with the cyclotron frequency $\omega^{i}_{B}$) and in this limit, the particle velocity
$v^{i}_{z} = v^{i}$.  The rate of change of the relative phase of O- and E-modes decreases with distance propagated as the plasma density falls and the polarization of the radiation approaches asymptotically what is observed. In terms of observer intensities, this is in general a sum of linearly and circularly polarized components. For the conditions assumed by Jones (2016), the unequal amplitudes of the O- and E-modes result in the linear component being not much changed from the original linear polarization of the emission region. Then the circularly polarized component is relatively small and its intensity and sense depend on the final relative phase of the two plasma normal modes. Along the line of sight, this is
\begin{eqnarray}
\phi_{ret} = \frac{\omega}{c}\int^{\infty}_{r_{e}}\Delta n_{O}dr = \sum_{i}
\frac{r_{e}\omega^{2}_{p}}{4\gamma_{i}\omega c}.
\end{eqnarray}
This is valid for small $r_{e}$ such as follow by acceptance of the Hassall et al estimates given in Table 1.  In this case $\theta_{k} \sim \gamma^{-1}_{i}$, as assumed in Jones (2016).  If flux-line curvature were thought to be relatively more significant, as it might be at emission radii an order of magnitude larger, $\Delta n_{O}$ would be reduced by a factor of $\gamma^{2}_{i}\theta^{2}_{k}$ but the phase retardation given by equation (6) by only a much smaller amount because most of the phase retardation occurs at radii below $2r_{e}$. 

A large part of the seminal paper by Cheng \& Ruderman (1979) on propagation and circular polarization was concerned with the effects of refraction in electron-positron plasmas, in particular, the conditions under which coherence between O- and E-modes would be maintained.  But in an ion-proton plasma, $\Delta n_{O}$ is so small that the angular deviation between O- and E-modes produced by refraction in gradients perpendicular to ${\bf B}$ is negligible as is the relative lateral displacement of the modes.

The final relative phase which determines the polarization state must be a function of observed longitude within the profile even if only as a consequence of the variation of particle Lorentz factor across the open magnetosphere. There is a change in the sign of $V$ (left or right circular polarization) for each increment of $\pi$ change in $\phi_{ret}$, interspersed by zeros in $V$. Examination of the longitude dependence and changes of sense of the circular polarized intensity in each of the Karastergiou \& Johnston profiles shows that these vary quite slowly, from which it follows that the absolute relative phase change $\phi_{ret}$ occurring after the surface of last absorption cannot be large but must be no more than of the order of $1 - 10 \pi$. This is qualitatively consistent with equation (6) for an ion-proton plasma and $r_{e}\sim 10^{2}$ km.  
  
Components of the dielectric tensor are inversely proportional to particle mass so that even a small electron-positron number density has an overwhelming effect on the refractive index of an ion-proton plasma given by equation (5).  A Goldreich-Julian current-density primary beam with $N_{\pm} \sim 10^{2}$ pairs per positron would produce phase retardation several orders of magnitude too large even at emission radii $r_{e} \sim 10^3$ km and its rate of variation would be much faster than is observed. The rapid changes of sense would render circular polarization unobservable.  This has to be viewed in the context of Table 1 on which basis we do not believe that the electron-positron plasma is a possible source.

There is a further factor in favour of the ion-proton plasma.  Owing to the rapid decrease of $\omega^{i}_{B}$ with altitude in both electron-positron and ion-proton plasmas, there are regions nearer the light cylinder in which cyclotron absorption occurs.  In the electron-positron case, the absence of any strong observational evidence for this process in pulsars has been considered in the past to be a very real problem (see Blandford \& Scharlemann 1976; Mikhailovskii et al 1982; Lyubarskii \& Petrova 1998; Fussell, Luo \& Melrose 2003).  The explanation is that the cross-section in the case of protons is negligibly small.

\section{Properties of the ion-proton plasma}

\subsection{Consequences of reverse electron flow}

A reverse flux of electrons from photo-ionization (or pair creation should it occur) must certainly be a feature of polar caps.  The consequent physical processes are all individually well-established and their application to polar caps was described by Jones (2010a, 2011).  A summary is given here.  Formation of the nuclear giant dipole state by electromagnetic shower photons in the energy interval $15 - 30$ Mev is the dominant hadronic interaction giving neutrons and protons. The photon track length in an electromagnetic shower per unit photon energy and primary electron energy is known, as are the cross-sections (integrated over photon energy) for giant dipole-state formation.  From these, the proton production rate is easily obtained. The polar-cap surface, whether liquid or solid, lies under a thin atmosphere of ions, in local thermodynamic equilibrium (LTE), having a scale height of about $10^{-1}$ cm.  Its column density is an approximately exponential function of surface temperature and ion work function, and therefore is difficult to estimate.  Electromagnetic showers are formed either partially or entirely within this atmosphere depending on its column density.  Protons emitted are quickly reduced to thermal energies and diffuse to the top of the atmosphere with negligible probability of nuclear interaction, and with characteristic time $\tau_{p} \sim 1$ s.

Atmospheric structure is important.  Given LTE, it depends on the charge-to-mass ratio of its components.  Thus the protons cannot be in static equilibrium within the predominantly ionic atmosphere.  They pass through and are preferentially accelerated from the polar cap.  Ions are accelerated only if there are insufficient protons to form a Goldreich-Julian current density.  The relation between ion and proton current densities is local and time-dependent.

With regard to the effect of high static magnetic fields on shower formation, approximate cross-sections for electron-positron pair creation and bremsstrahlung have been obtained up to $B \sim B_{c} = 4.41\times10^{13}$ G (Jones 2010b).  The overall influence of high fields on shower formation is complex. The Landau-Pomeranchuk-Migdal effect (see Klein 1999) is present. It represents the effect of small-angle Coulomb multiple scattering in the medium in partially removing the coherence intrinsic to the Feynman diagram description of these processes, which becomes significant at high electron energies and at high matter densities , as at the neutron-star surface. It reduces bremsstrahlung and pair-creation cross-sections significantly at $B_{c}$.  But an order of magnitude below this, the conclusion, adequate for present purposes, is that proton creation per GeV shower energy is only a weakly varying function of $B$ and in the interval of interest here does not differ much from its zero-field value.

It was recognized in Jones (2012a, 2012b) that the presence of reverse electrons from photo-ionization adjusts the charge density above the polar cap so as to limit the acceleration field ${\bf E}_{\parallel}$ to values much below those predicted by the Lense-Thirring effect (Beskin 1990, Muslimov \& Tsygan 1992).  The effective blackbody field is not that of the polar cap but of the surface area, temperature $T_{s}$, of about a steradian centred on the polar cap, whose photons transformed to the ion rest frame can reach the photo-ionization threshold at those altitudes where the Lense-Thirring acceleration field remains significant.  The number of photons reaching this threshold increases rapidly with increasing ion Lorentz factor with the consequence that ion acceleration is self-limiting.   At temperatures $T_{s} > 1-2\times 10^{5}$ K, no more than moderately relativistic Lorentz factors are reached: ${\bf E}_{\parallel}$ is limited to values far below those needed for pair creation.
A model polar cap was constructed on these bases in Jones (2013a).

But the problem with these papers is that, whilst the processes were in essence correctly investigated, their application to the observed classes of pulsar including the Rotating Radio Transients (RRAT) was flawed owing to the assumption that a secondary electron-positron plasma could be a source of the coherent emission.  The properties of  the ion-proton plasma above the polar cap as they appear in the Jones (2013a) model can be summarized as follows.

(i)		The effect of photo-ionization is local.  The proton flux at a point on the polar cap is determined exclusively by the previous reverse-electron energy flux at that point.

(ii)	The state of the whole polar cap is chaotic: the rate of change of ion or proton current densities is characterized by $\tau_{p}$.  Thus the fluctuations in ion and proton current densities are broadly consistent with the intensity differences observed between successive individual pulses.

(iii)	Coincident non-zero ion and proton current densities in any part of the polar cap are necessary for Langmuir-mode growth above that area.  Absence of either component results in no coherent emission except that very weak emission may be observed if there is a small background of electrons or positrons in addition to the baryonic component (see Jones 2014c).

(iv)	For a certain interval of $T_{s}$, mixed current densities occur preferentially near the polar-cap perimeter, with central regions having almost entirely proton current densities for much of the time.

(v)		The ion Lorentz factors reached are approximately those for which the blackbody photons in the ion rest frame reach successive photo-ionization thresholds.
Hence, very roughly, $\gamma_{i} T_{s}$ can be regarded as a constant.

(vi)	Low values, roughly $T_{s} < 2\times 10^{5}$ K, are insufficient for photo-ionization unless Lorentz factors are high.  The Langmuir mode growth-rate is
$\exp \Lambda$ where $\Lambda \propto \gamma^{-3/2}_{i}$, so that in the low-$T_{s}$ and presumably old-age limit, growth to a turbulent state does not occur.  The neutron star is not then observable as a radio pulsar.

\subsection{Multi-component Langmuir modes and bi-stability}

The growth rate $\exp \Lambda$ for either longitudinal or for the quasi-longitudinal Langmuir modes described by Asseo, Pelletier \& Sol (1990) are large for an ion-proton plasma of moderate Lorentz factor.  Approximate rates were given by Jones (2012a, 2012b).  The conditions for growth are ideal in that the velocity distributions for each component are essentially $\delta$-functions and the growth rate is not strongly dependent on the ratio of the two current densities. Growth rates are such that an exponent $\Lambda = 30$ can be reached at altitudes of $\sim 2R$.

The dielectric tensor component $D_{zz}$ in the high-field limit (the $z$-axis is locally anti-parallel with ${\bf B}$) for an ion-proton plasma with the addition of a continuum electron-positron component is,
\begin{eqnarray}
D_{zz} = 1 - \frac{\omega^{2}_{p1}}{\gamma^{3}_{1}(\omega - k_{\parallel}v_{1})^{2}}
 - \frac{\omega^{2}_{p2}}{\gamma^{3}_{2}(\omega - k_{\parallel}v_{2})^{2}} \nonumber  \\
  + \frac{m\omega^{2}_{pe}}{k_{\parallel}}\int^{\infty}_{-\infty}dp \frac{\partial f}{\partial p}\frac{1}{\omega - k_{\parallel}v(p) + i\epsilon}.
\end{eqnarray}
Here $k_{\parallel}$ is the wave-vector component anti-parallel with ${\bf B}$, $m$ is the electron mass and $f(p)$ its momentum distribution normalized to unity.  The indices $1$ and $2$ here refer to ions and protons, respectively, and
$\omega^{2}_{i} = 4\pi n_{i}q^{2}_{i}/m_{i}$ in which $n_{i}$ and $q_{i}$ are the mass and charge of particle $i$.  In the high-field limit, the velocities are all anti-parallel with the local ${\bf B}$.  The final term in equation (7) is identical with the expression given by Buschauer \& Benford (1977).  The Langmuir modes are defined by $D_{zz} = 0$ and its evaluation for an electron-positron distribution uniform between limits $p_{0}$ and $p_{m}$ shows that, dependent on $\gamma_{0}$ and $\gamma_{m}$, $D_{zz}$ is dominated by the electron-positron component even at number densities one or two orders of magnitude smaller than Goldreich-Julian.  For $\gamma_{0,m} > \gamma_{1,2}$, ion-proton Langmuir modes exist (see Jones 2014c) but for $\gamma_{0} < \gamma_{1,2}$, the mode has no growth rate except at number densities two orders of magnitude smaller than Goldreich-Julian.

It is very difficult to be precise about the sources of small electron-positron number density components at altitudes below $r_{e}$ or about their Lorentz factors in the emission region.  Apart from the usual magnetospheric processes, there is a small flux of backward-moving MeV photons from the photo-electron showers.  Its sources are successive Compton scattering within the shower and the nuclear capture of neutrons produced by giant dipole-state decay.  The magnetic conversion of these photons is very much a function of the polar-cap magnetic field. One reason why their flux can change is that the atomic numbers of ions in the LTE atmosphere in the region of the showers is subject to the medium time-scale instability described by Jones (2011). This is a consequence of the gradual change in atomic number of nuclei within the shower region as they move towards the surface, caused by continual giant dipole-state formation. Thus the surface $Z$ is much below the initial $Z_{0}$ of the undisturbed surface elsewhere on the star. The order of magnitude of the associated time is that for the emission at Goldreich-Julian current density of ions equivalent to one radiation length,
\begin{eqnarray}
\tau_{rl} \approx 2.1\times 10^{5}PZ^{-1}B^{-1}_{12}\left(\ln\left(12Z^{1/2}B^{-1/2}_{12}\right)\right)^{-1} \hspace{3mm} s,
\end{eqnarray}
which is also of similar order of magnitude to mode-change and null time-scales. Hence
the capacity of even a small electron-positron number density to reduce ion-proton Langmuir-mode growth rates to zero is of direct interest in relation to the magnetospheric bi-stability which is apparent in the mode-changes and nulls that are observed in some pulsars.  This was considered by Jones (2015a) and here we refer specifically to Fig. 1 of that paper.  The Langmuir-mode growth rate exponent is $\Lambda \propto B^{1/2}_{12}P^{-1/2}$ and Fig. 1  gives this quantity as a function of $B$, extending over seven orders of magnitude, for various classes of neutron star or pulsar.  It is immediately obvious that the existence of mode-changes and nulls is a function of $B$.  There appears to be no evidence of them in the MSP.  They begin to be present at $B\sim 10^{11}$ G and appear more frequently with increasing $B$ until the RRAT are reached at $B\sim 10^{13}$ G.  To maintain the clarity of the diagram, it does not show the very large number of normal pulsars in which mode-changes or nulls are not observed.  But their average value of $B$ obtained from the ATNF catalogue (Manchester et al 2005) is lower than for pulsars with nulls.  Radio emission described as primary is not usually observed at $B\sim 10^{14}$ G.  The conclusion of Jones (2015a) was that this sequence simply reflects the fact that the rate of background pair creation by single-photon magnetic conversion from the photon sources discussed earlier in this Section is a very strongly varying function of $B$.  Hence it is negligible in the MSP but becomes progressively more likely as $B$ increases.  In the intermediate-$B$ region, changes in the background electron-positron density arising from changes in the surface atomic number $Z$ are sufficient to lead to changes in Langmuir-mode growth rate over the whole or some part of the polar cap, with time-scales of the order of $\tau_{rl}$. Therefore, mode-changes and nulls are expected to be a feature of intermediate values of $B$, as is observed.  High electron-positron background densities at large $B$ reduce growth rates to zero so that no primary radio emission is observed.

\subsection{Further problems}

The set of profiles published by Karastergiou \& Johnston are of such complexity that there is some difficulty in accepting that they are not stochastic but are integrated and long-term stable.  This complexity must be anchored by some fixed feature of the polar cap.  If the surface were solid, inhomogeneity in $Z$ with a transverse length-scale of the order of $10^{3}$  cm would be one possibility, but there appears to be no reason to expect this.  A second and perhaps more plausible possibility is that the shape of the open magnetic flux lines is complex and possibly age-dependent at the surface, but approaches dipole form at $r > r_{e}$.  This kind of structure, specific to an individual neutron star, would imply that realistically, there may be limits to what can be understood in detail.  Whilst a physical mechanism for magnetospheric bi-stability has been identified in this work, its detailed application to mode-changes and nulls in individual pulsars may prove impossible.

The same comment can be made with regard to subpulse drift.  This was considered by Jones (2014a) where it was stated that there is little evidence for a systematic rotation pattern of plasma about the magnetic axis caused by ${\bf E}\times{\bf B}$ drift.  Instead, a repetitive travelling modulation of the ion and proton current densities was proposed along arcs on the polar cap.  This would permit the changes of drift direction that are observed.

The actual nature of the black-box interior is also a problem.  Growth of either longitudinal or quasi-longitudinal Langmuir-mode amplitudes leads to non-linearity and, it is reasonable to expect, the development of turbulence, also the consequent movement of energy to higher wavenumbers.  But in Jones (2016), no more than order-of-magnitude estimates of time-scales were given.  A much more complete analysis of turbulence in the high-field limit and its coupling with the radiation field is needed.

The complete set of papers summarized here assumes corotation in the closed sector of the inner magnetosphere as in the paper of Goldreich \& Julian (1969) in which the induction field of a non-aligned neutron star was neglected. Melrose \& Yuen (2012, 2014) have shown recently that its inclusion is likely to lead to a departure from corotation.  This has no essential effect on the conclusions of these papers (see Jones 2015b) but indicates that in detail, the structure of polar-cap current densities may be more complicated than we have assumed. There may be detail that is within the capability of new observing instruments such as the Square Kilometre Array (SKA).

\section{Implications}

Although there are uncertainties of detail, we believe that there is good observational evidence for the existence of an ion-proton plasma, with the possibility of at most a very small background of electrons and positrons, in radio-loud pulsars. This defines the plasma condition in the inner magnetosphere with some certainty for the ${\bf \Omega}\cdot{\bf B} < 0$ case and has implications for the further study of neutron stars which we wish to outline here.

The application of computational techniques in numerical plasma physics to the pulsar magnetosphere (see, for example, Bai \& Spitkovsky 2010) assumes, following Mestel et al (1985) and Beloborodov (2008), that the open magnetosphere current density is determined by the state of the entire system and, crucially, that this is well-approximated by the outer magnetosphere.  Thus complete solutions are obtained external to a sphere of radius $\sim 0.2R_{LC}$, usually under the assumption of force-free electrodynamics.  Particles of either sign can pass freely through this surface and their charge and current densities ($\rho$ and
${\bf j}$)  in certain regions of the solution satisfy $|{\bf j}| > |\rho c|$, which is possible only with particle counterflow requiring pair creation. The same assumption is the basis for the very recent computational paper of Philippov et al (2015) on the Lense-Thirring effect.  These authors assume that the open-magnetosphere current density is set in the vicinity of the light cylinder so that near the neutron-star surface, its magnitude exceeds the corotational Goldreich-Julian charge density as modified by the Lense-Thirring effect.  (This is the opposite effect as compared with, for example, Harding \& Muslimov 2002, in whose work it is lower) In the work of Philippov et al, there is also electron acceleration to ultra-relativistic energies with the concomitant formation of a  dense secondary electron-positron plasma.

As stated previously (Jones 2013a), we do not accept this view and regard the high-field and high energy-density region of the polar cap as setting what amount to boundary conditions for the true outer-region solution.  Interpretation of the circular polarization observations with only minimal assumptions does not support the presence of such a plasma in radio-loud pulsars and there are other reasons set out in Section 2 why such a source of emission would not match observations.  These arguments, we believe, show that the basis for their computations is incorrect, at least for ${\bf \Omega}\cdot{\bf B} < 0$ neutron stars.

A possible alternative approach to computational solutions for the outer magnetosphere would be, in outline, to accept the properties found for the inner magnetosphere as forming boundary conditions for the outer solution. In the ${\bf \Omega}\cdot{\bf B} < 0$ case, the conditions would be those of the ion-proton plasma.  The case of ${\bf\Omega}\cdot{\bf B} > 0$ neutron stars is different because protons produced in reverse-positron showers have no effect.  But there appears to be no reason why the  current density should not be determined at the polar cap, as in the ${\bf \Omega}\cdot{\bf B} < 0$ case.

Involvement of the neutron-star surface as a condensed-matter system with additional degrees of freedom as well as an electromagnetic boundary condition means that the ${\bf \Omega}\cdot{\bf B} < 0$ case is very rich in possible modes of behaviour as compared with ${\bf \Omega}\cdot{\bf B} > 0$.  In this respect alone it appears a very much better match for the observed range of pulsar phenomena. The drawing of a clear distinction between ${\bf \Omega}\cdot{\bf B} > 0$ and
${\bf \Omega}\cdot{\bf B} < 0$  neutron stars and the identification of the latter with normal radio-loud pulsars and MSP raises the question of how the former are observed in the electromagnetic spectrum.  Insofar as incoherent emissions are concerned there should be little difference provided, as is likely, the ultra-relativistic electrons or positrons necessary for these processes are present near the light cylinder.  Secondary pairs can be produced by curvature radiation in small-$P$ neutron stars or more generally in limited densities by inverse Compton scattering at the polar cap (see Hibschman \& Arons 2001: Harding \& Muslimov 2002). But it may be that the conditions necessary for pair creation in the outer magnetosphere can be determined with certainty only by inspection of numerical solutions.  It is also unclear whether or not the electron-positron number densities present in a secondary plasma are needed or play any significant part in incoherent emission such as high-energy $\gamma$-rays. There is some magnetar radio emission (for example, the emission of J1810-197 has a small negative spectral index to 144 GHz: Camilo et al 2007, see also Serylak et al 2009) and such emission could be evidence for the ${\Omega}\cdot{\bf B} > 0$ class.

\section*{Acknowledgments}

I thank the anonymous referee whose reading and comments have much improved the presentation of this work.

\bsp

\label{lastpage}

\end{document}